\documentclass{article}

\usepackage[preprint]{nips_2018}

\usepackage[utf8]{inputenc} 
\usepackage[T1]{fontenc}    
\usepackage{hyperref}       
\usepackage{url}            
\usepackage{booktabs}       
\usepackage{amsfonts}       
\usepackage{nicefrac}       
\usepackage{microtype}      
\usepackage{graphicx}
\usepackage{subfigure}
\usepackage{natbib}

\begin{document}
\title{Data-parallel distributed training of very large models beyond GPU capacity}

%

\author{
  Samuel Matzek \\
  IBM \\
  \texttt{smatzek@us.ibm.com} \\
  \And
  Max Grossman \\
  BP \\
  \texttt{max.grossman@bp.com} \\
  \And
  Minsik Cho \\
  IBM Research\\
  \texttt{minsikcho@us.ibm.com} \\
  \And
  Anar Yusifov \\
  BP \\
  \texttt{anar.yusifov@bp.com} \\
  \And
  Bryant Nelson \\
  IBM \\
  \texttt{bryant.nelson@ibm.com} \\
  \And
  Amit Juneja \\
  IBM \\
  \texttt{amit.juneja@ibm.com} \\
}
\maketitle

\begin{abstract}
GPUs have limited memory and it is difficult to train wide and/or deep models that cause the training process to go out of memory. It is shown in this paper how an open source tool called Large Model Support (LMS) can utilize a high bandwidth NVLink connection between CPUs and GPUs to accomplish training of deep convolutional networks. LMS performs tensor swapping between CPU memory and GPU memory such that only a minimal number of tensors required in a training step are kept in the GPU memory. It is also shown how LMS can be combined with an MPI based distributed deep learning module to train models in a data-parallel fashion across multiple GPUs, such that each GPU is utilizing the CPU memory for tensor swapping. The hardware architecture that enables the high bandwidth GPU link with the CPU is discussed as well as the associated set of software tools that are available as the PowerAI package.

\end{abstract}
\section{Introduction}
Deep convolutional neurals networks (CNNs) have been shown to be highly effective in tasks such as image segmentation for both 2-D and 3-D images. Training for CNNs is typically done on graphical processing units (GPUs) while utilizing thousands of parallel cores for weight update algorithms such as stochastic gradient descent (SGD). However, GPUs have a small memory space (12-32GB), and, in contrast, CPUs use a far more scalable type of DRAM memory (DDR3 or DDR4) and can have 256-2048GB memory capacity. GPU memory capacity has been relatively constant for the past 2-3 generations, while deep neural network models are getting deeper and wider to achieve higher learning capacity, for example, \cite{resnet1001} proposes a Resnet with 1001 layers. Therefore, a complex neural network which would be perfectly trained on CPUs cannot be trained on GPUs due to the limited device memory.  

Large Model Support (LMS) \cite{tflms} was developed for using both the CPU memory and the GPU memory together, in order to enable deep learning to continue to push boundaires. LMS depends on swapping tensors between CPUs and GPUs, and therefore, taking advantage of a fast connection (aggregated bandwidth of 300GB/s) between the two types of computing units. Such a fast link was developed by IBM's incorporation of NVIDIA's NVLink communications protocol in a series of IBM Power Systems servers. In this paper, a brief review is provided for LMS and IBM Power Systems, and experiments are presented for 3-D image segmentation tasks that show how these technologies can be utilized for practical deep learning tasks. Not only is it critical to be able to train deeper models while utilizing the CPU memory, it is equally important to be able to train on a large number of data samples in a data-parallel way. It is shown in the paper how LMS can be applied in conjunction with an MPI based distributed training framework to utilize multiple GPUs for data-parallel training. Experiments from early adopters of the technologies are also included to show the effectiveness of LMS and DDL on Power Systems for training deep convolutional networks.

\section{Power Systems and PowerAI}
The IBM Power System AC922 is the latest generation of the IBM POWER9 processor-based systems designed for high performance deep learning and artificial intelligence (AI). In partnership with NVIDIA and other OpenPOWER ecosystem members these systems form the \#1 Top 500 DOE Summit supercomputer. Performance was in part driven by the POWER9 CPU and the NVLink 2.0 connection between CPUs and GPUs. In non-POWER based platforms NVLink is only used to interconnect GPUs. The AC922 system has: two POWER9 processors (40 or 44 cores); 1 TB DDR4 per socket for a maximum of 2 TB memory; 4 or 6 NVIDIA Tesla V100 GPUs (7.8 teraflops at double precision); and interconnects for NVLINK 2.0, PCIe Gen4, CAPI 2.0, and OpenCAPI that provide 2X to 5.6X bandwidth w.r.t. PCIe Gen3. In this work, deep learning experiments are performed on AC922 servers and previous generation POWER8 systems (NVLink 1.0 GPU-CPU connections). 

PowerAI is a software distribution of popular open source deep learning frameworks (TensorFlow, Caffe, PyTorch, etc.) that are co-optimized for POWER CPU based servers using NVLINK. PowerAI includes the tools Distributed Deep Learning (DDL) and LMS. DDL enables training of deep learning networks on multiple GPUs across multiple server nodes; achieving near linear scaling at 95\% efficiency on popular data sets. LMS allows memory swapping of tensors with CPU memory so that deep networks that don't otherwise fit into the GPU memory can be trained with little overhead. Note the experiments in this work are conducted with TensorFlow however LMS and DDL capabilities are also available in PowerAI for Caffe, and support for PyTorch is being added.

\subsection {PowerAI DDL}
In this section, we discuss an efficient MPI-based communication library (PowerAI DDL) for deep learning that provides
a highly efficient all-reduce algorithm for SGD, and we report its performance.
All-reduce operation is  an integral part in modern deep learning frameworks~\cite{chainer,cntk,mxnet,caffe2,jia2014caffe,tf,fb256,nccl,fb_gloo} due to
its efficiency over a parameter-server based approach~\cite{federated_training,coop_sgd}.
The key concept in DDL is to decompose one all-reduce operation into a series of 
reduce-scatter and all-gather patterns in a topology-aware fashion. This  enables a large-scale deep learning approach with reduced communication overhead.
The key features in DDL can be summarized as follows: (1) DDL adapts to the hierarchy of communication bandwidths by leveraging topology-awareness, so that it fully utilizes the heterogeneous network architecture in popular deep learning platforms~\cite{DGX1,minsky}, (2) Through topology-aware decomposition, DDL also minimizes the  communication latency overhead, the critical bottleneck in large-scale deep learning and (3)  For each decomposed piece, DDL can mix-and-match various reduce-scatter and all-gather implementations/algorithms over different network fabrics to maximize network utilization.

\begin{figure}[!t]
	\begin{center}
		\begin{tabular}{ll}
			\subfigure[\small{Intel systems with PCIe gen3 and 10Gbps Ethernet}]{	\includegraphics*[width=2.5in]{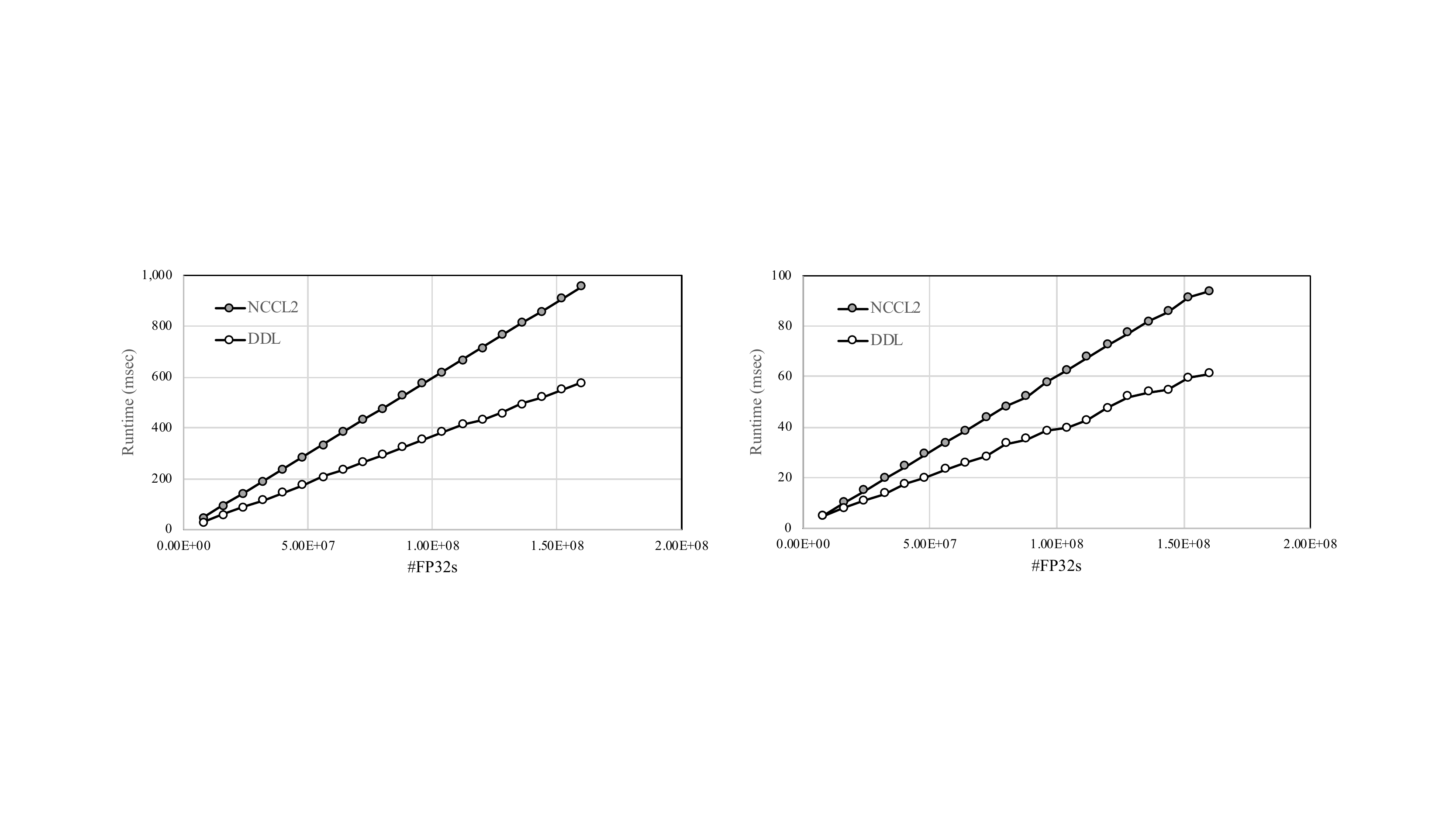}}
			&
			\subfigure[\small{IBM systems with NVLink and 100Gbps InfiniBand}]{	\includegraphics*[width=2.5in]{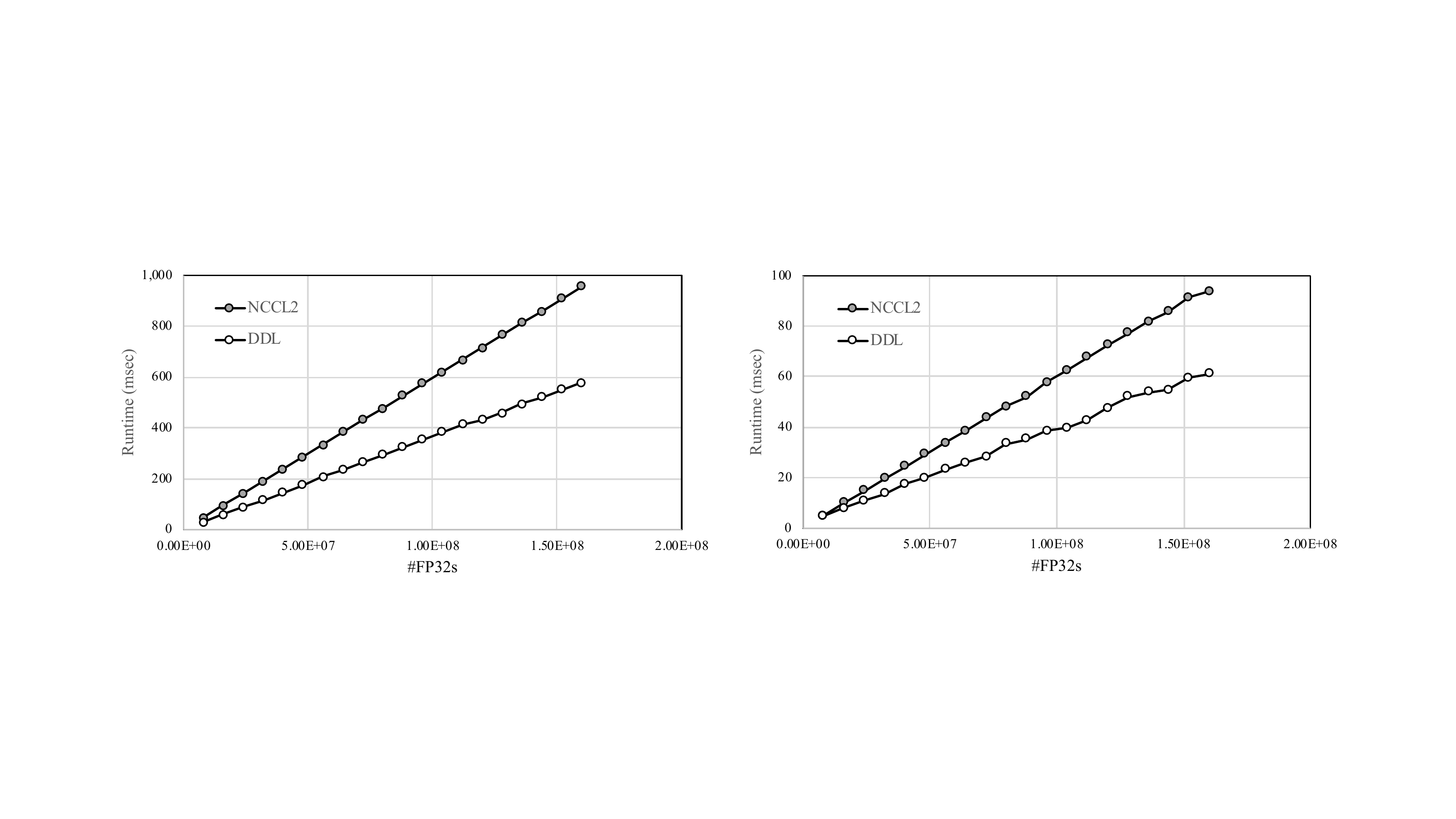}}
		\end{tabular}
		\caption{\texttt{All-reduce} performance on two systems.}
		\label{blc:all}
	\end{center}
\end{figure}

PowerAI DDL is easy to apply to existing code. For example, in TensorFlow, DDL can be achieved by adding the "import ddl" line, using an MPI-like "rank" function to specify how data is split across GPUs, and running python code through a "ddlrun" command. We report the pure \texttt{all-reduce} performance comparison between DDL and NCCL (i.e., ncclAllReduce) on two different setups.
In one setup, we used two Intel Xeon(R) CPU E5-2680 systems with 4 Nvidia Telsa P100-PCIE-16GB GPUs each, connected through 10Gbps Ethernet. Within the Intel systems,
the GPUs are connected through PCIe gen3.
In the other setup, we used two  IBM Power8 S822LC systems (previous generation from AC922) with 4 NVidia Tesla P100-SXM2 GPUs each, connected through 100Gbps InfiniBand. Within the IBM systems, the GPUs
are connected through NVLink ~\cite{minsky,nvlink}.
Figure ~\ref{blc:all} shows that DDL outperforms NCCL by 1.6X, exploiting the network hierarchy within systems as well as between systems on both setups, over a wide range of 
FP32 floating-point number counts. 

\subsection{Large Model Support}

When TensorFlow trains a neural network, the model, training data, and neural network operation executable binaries are loaded into the GPU memory. While the operations in the neural network run, they produce feature or parameter maps as output. These feature maps are called tensors and usually take the form of multi-dimensional array data. During training these tensors will also reside in GPU memory until no remaining operation needs the data, at which point they are garbage collected. In many models some tensors can be long lived, and these long-lived tensors can also be the largest of the model. For example, the tensors produced by the first few layers of a CNN are typically the largest and remain in memory until the last stages of backward propagation. LMS reduces the amount of memory used by the GPU during model training by swapping these tensors out to system memory. This allows downstream operations to have room in GPU memory for their tensors. The swapped out tensors can be swapped back into GPU memory when they are needed as input to later operations.

TensorFlow allows the specification of which computational device (CPUs, GPUs, etc) will run a given operation in the model graph. The swapping of tensors during neural network execution has been implemented by analyzing the graph before training or inferencing and inserting operations that do not change the tensors, but rather trigger the tensors to move between GPU and CPU memory. This can be done by inserting "Identity" operations into the graph between producing and consuming operations of tensors on the GPU. The identity operations are specified to run on the CPU which implicitly trigger the swap out of the tensor. Since the inputs and outputs of the original operation are re-mapped to go through the swap-out and swap-in operations, the tensors will move to system memory and then back into GPU memory for the original consuming operation. Further information on the tensor swapping methods can be found in \cite{tflms}. A Python module named TensorFlow Large Model Support (TFLMS) was produced that does the static graph analysis and inserts the swapping nodes to temporarily move tensors to system memory during graph execution. 

\section{Image Segmentation with LMS and DDL}
\begin{figure}
	\begin{center}
	\begin{tabular}{ll}
		\subfigure[]{\includegraphics[width=2.9in]{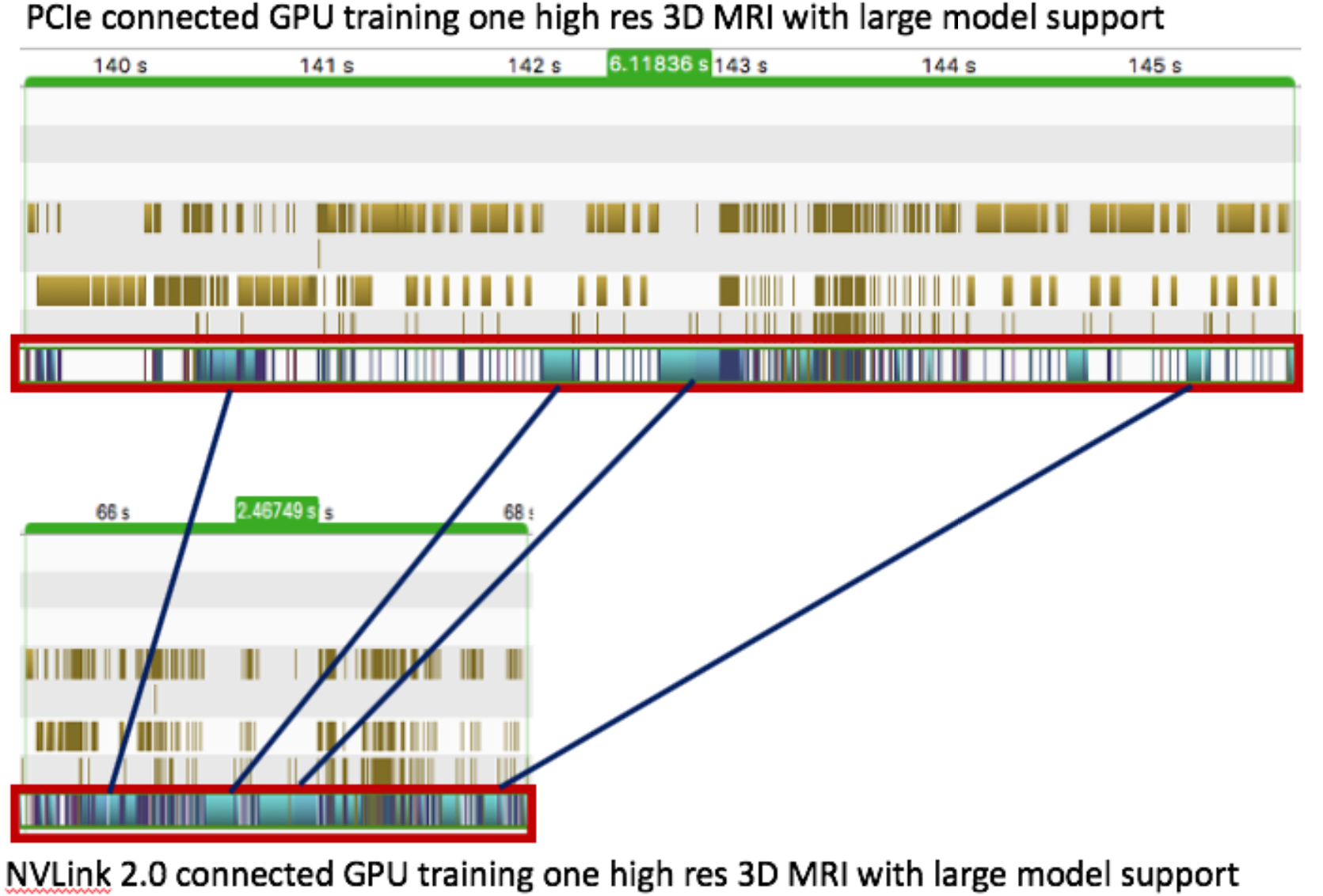}}
			&
		    \subfigure[]{\includegraphics[width=2.5in]{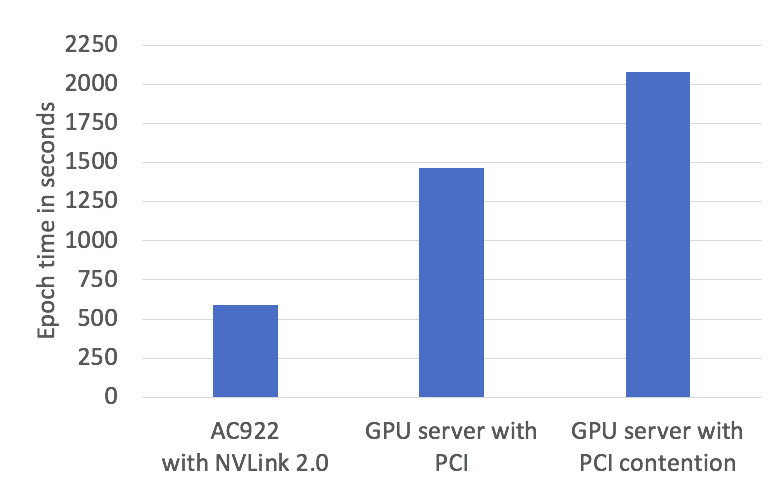}}
	\end{tabular}
		\caption{Performance comparison on NVLink and PCIe connections to GPU: (a) Data transfer/computation profile on servers with NVLink vs PCIe connections to GPU, (b) Training time on servers with NVLink and PCIe connections to GPU}
    \label{fig:nvprof}
	\end{center}
\end{figure}

To demonstrate TFLMS in a real-world use case we use a Keras model, ~\cite{dellis}, written to process the TCGA and MICCAI BraTS 2017 datasets ~\cite{brats,brats2,bratsdata1}. In this work the model code was modified to work with the Keras APIs included in TensorFlow 1.8 and PowerAI 1.5.2. The model could be trained at a maximum resolution of $144^3$ without failing on memory errors on the NVIDIA Volta V100 GPU with 16GB of memory. When using TFLMS to swap tensors it could be trained on images at a resolution of $192^3$, a 2.4x increase in resolution. To investigate the effect of the POWER9 NVLink 2.0 connections between the CPU and GPU, and its higher speed memory bus, the model was trained at the maximum $192^3$ resolution on a POWER9 server with NVIDIA Volta V100 GPUs. It was also trained on a server that had pairs of NVIDIA Volta V100 GPUs connected to PCIe Generation 3 buses in such a way that two GPUs would share a single PCI bus. The training runs were profiled with the NVIDIA profiler, nvprof, and the profile was then analyzed in NVIDIA Visual Profiler. A comparison of training on a single MRI image through the model on both servers can be seen in Figure \ref{fig:nvprof}(a).  It is immediately evident that the time series for the AC922 is shorter than the PCI connected GPU server. The processing of a single MRI by the model takes 2.47X longer on the PCI connected GPU than on the GPU connected by NVLink. The utilization of each of the GPUs is bound by a red rectangle. The white space on the GPU utilization line denotes when the GPU compute utilization is 0\%. 

The PCIe connected GPU compute utilization goes idle often, and this is a direct result of the memory copy due to swapping and graph execution waiting for operation inputs to arrive before proceeding. As previously noted, since the GPUs on the two systems (with and without NVLink) are the same model and so corresponding operations in the graph take the same amount of time in each. Some corresponding operations between the two runs are connected with lines. When multiple GPUs are used for training and two GPUs share a PCI bus, the contention on the bus increases the training time further. Figure \ref{fig:nvprof}(b) shows the training time goes from almost 2.5x slower to 3.5x slower than the NVLink 2.0 connected GPU. Therefore, the NVLink 2.0 connected GPUs in the POWER9 server allow much faster training when using TFLMS. To measure the overhead of TFLMS on model training, the resolutions between $144^3$ and $192^3$ were tuned and trained with their optimal TFLMS parameter settings on an AC922 with 16GB NVIDIA Tesla V100 GPUs. The model resolutions were then trained without using TFLMS on an AC922 with 32GB NVIDIA Tesla V100 GPUs. This allows us to measure the overhead of TFLMS separate from the overhead of dealing with larger data resolutions. The overhead percentages over the resolution factor above $144^3$ range from 3\% at 1.4X the resolution to 25\% at 2.4x the resolution.

\subsection{Putting LMS and DDL together}
With the increase in image resolution, the training time went up significantly so DDL was applied to distribute the training of the 3DUnet model.  To test the DDL integration the model was trained on 4 IBM AC922s with 100Gb/s Mellanox CX5 InfiniBand adapters connected via a 100Gb/s Mellanox SB7700 InfiniBand switch. The $192^3$ resolution has a training time of 590 seconds per epoch on a single GPU. With DDL, this training time becomes 150 seconds per epoch on a single node with 4 GPUs, 76 seconds per epoch on 8 GPUs across 2 nodes, and 40 seconds per epoch on 16 GPUs across 4 nodes. The number of epochs for the model to converge, loss, validation loss, and Dice coefficients of the models trained with DDL were equivalent to the models trained without DDL.
\begin{figure}[h]
    \centering
    \includegraphics[scale=0.4]{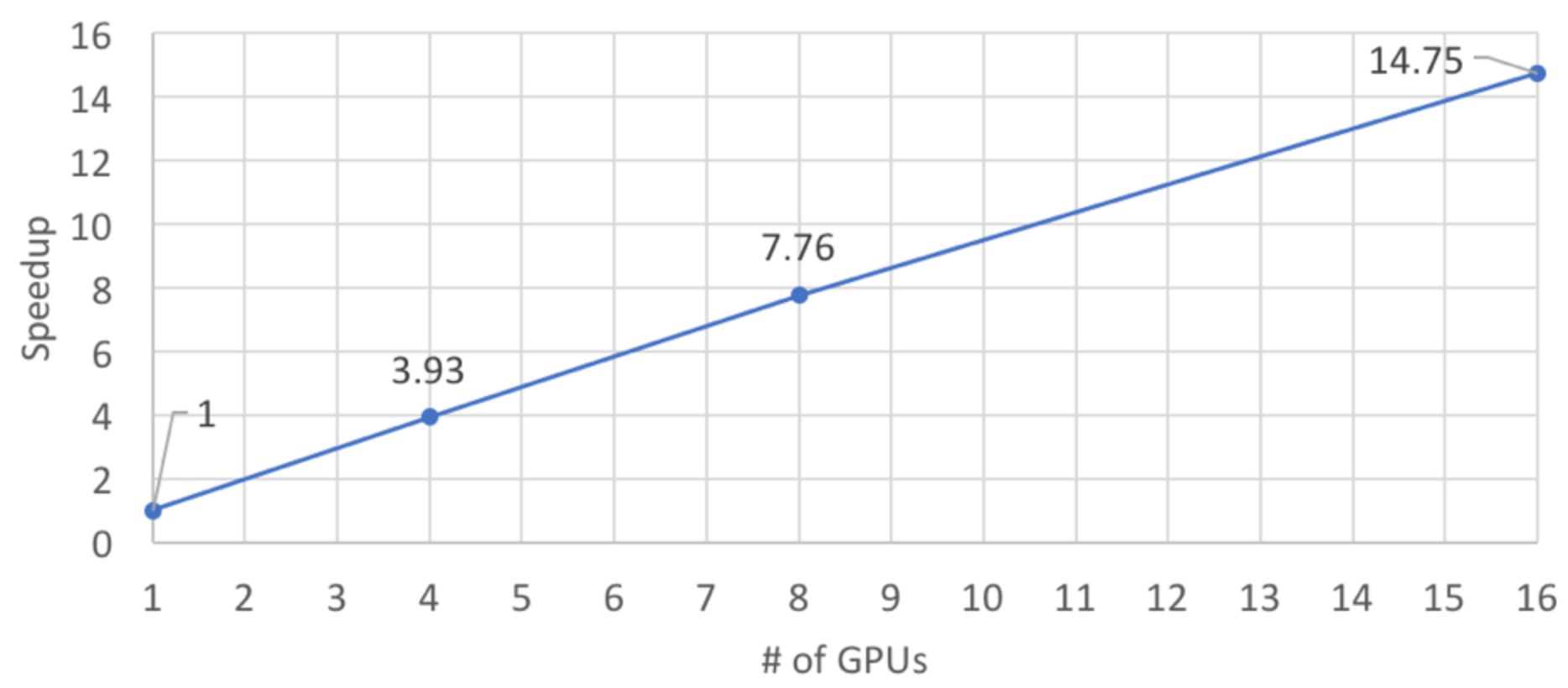}
    \caption{Scaling when using DDL with LMS}
    \label{fig:lmsddl}
\end{figure}

\section{End User Experience}
One of the early end users of DDL was the High Performance Computing team in
BP Inc.'s Center for High Performance Computing (CHPC).
BP's CHPC is responsible for supporting seismic processing and other
computationally heavy workloads using high performance computational resources,
which includes a small 12-node cluster of IBM Power8 nodes. Each Power8 node
includes two 10-core POWER8 processors and 4 NVIDIA P100 GPUs, all connected by
NVlink and with 1 TB of DRAM. Nodes are connected with an 100Gbps Infiniband interconnect.

BP's deep learning workloads today focus in 3D computer vision, unlike much of
the deep learning community which focuses more in the 2D space. Moving from 2D
to 3D kernels naturally yields an increase in throughput requirements, making
scale out more important. Additionally, 3D kernels and 3D data require more
memory to store, leading to a natural synergy with DDL and LMS. These combination of factors led BP's HPC team to evaluate the
performance and accuracy characteristics of both DDL and LMS. This evaluation took the form of porting an existing 3D Tensorflow model from
vanilla Tensorflow to DDL's APIs in two stages. First, the same model was
evaluated for scalability and convergence using distributed DDL jobs. Second,
LMS extensions were added to the code base and performance/convergence were
re-evaluated. This section offers: (1) A description of the experimental set up of the BP HPC team's tests. (2) A summary of the observed scalability (measured as time per epoch) and convergence of DDL. (3) A summary of the observed overheads and accuracy improvements with LMS. (4) A brief commentary on the programmability and long term maintenance impacts of using DDL.

\subsection{Experimental Setup}
The model used in these experiments is a 3D encoder-decoder convolutional model
for segmenting three-dimensional images. The input to the model is a
64$\times$64$\times$64 cube of voxels. This input first passes through two encoder
layers each consisting of a 3$\times$3$\times$3 convolutional layer followed
by a 2$\times$2$\times$2 max pooling layer, and outputting 128 channels per
voxel. Then, two inverse decoding layers follow, each with a 3$\times$3$\times$3 convolutional layer and an upsampling layer. The output for each
voxel is a likelihood measure for each of three possible classifications. The
final classification emitted is simply the one with the maximum likelihood.  
For training runs with vanilla Tensorflow or DDL, batches of 8 input blocks are used.  
For LMS experiments, the input block size was simply scaled up to
96$\times$96$\times$96 and layer sizes were adjusted accordingly. However, no
layers were added, nor were the attributes of any layers changed (e.g. pooling
factor). 

All experiments are run on the 12-node Power8 cluster using
the Univa GridEngine job scheduler and with exclusive access to all nodes.
In all experiments, a training dataset of 12.752 billion pre-labeled voxels was
used. Test accuracy is evaluated on $\sim$2.569 million voxels. This dataset
exhibits a class balancing problem -- 24.9\% of the test dataset is in class 0,
7.2\% is in class 1, and 67.9\% is in class 2. The weight of voxels are adjusted
during training to account for this. When doing multi-GPU training, the training
dataset is partitioned across GPUs (rather than replicated).

\subsection{Comparing Vanilla Tensorflow and DDL}

\begin{figure}[h]
    \centering
    \includegraphics[scale=0.4]{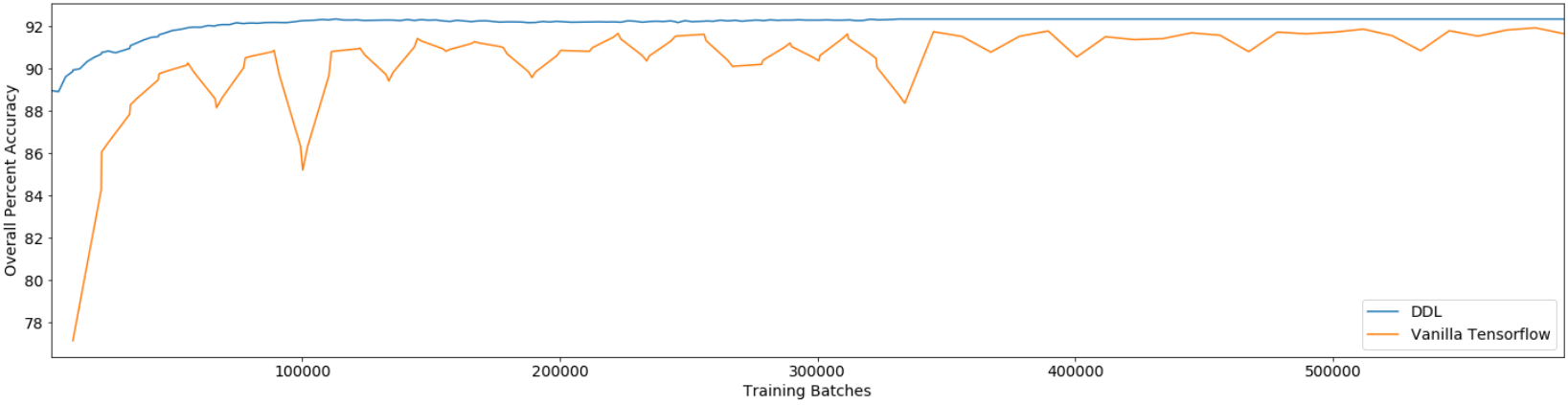}
    \caption{Convergence of DDL and Vanilla Tensorflow accuracy.}
    \label{fig:ddlconvergence}
\end{figure}

In comparing Vanilla Tensorflow and DDL, we focused on (1) the scalability of
DDL as more GPUs are added, and (2) how training runs converged in terms of accuracy.  
To measure scalability, we measure the wallclock time to complete a single epoch
for a varying number of GPUs. Table ~\ref{tab:ddlscaling} contains the results,
and shows that DDL training time drops nearly linearly with the number of GPUs
used. Figure~\ref{fig:ddlconvergence} plots the accuracy of the model as training
epochs progress on both Vanilla Tensorflow and DDL using the same optimizer and
learning rate. The DDL curves illustrate both smoother convergence and a higher
eventual accuracy for our model that works on 64$\times$64$\times$64 input (without LMS).

\begin{table}
\centering
\begin{tabular}{| c | c | c | c | c |}
\hline
	\textbf{\# GPUs} & \textbf{\# Nodes} & \textbf{Execution Time (s)}   & \textbf{Speedup w.r.t. Previous} & \textbf{\% Scaling w.r.t. 1 GPU}     \\\hline
	1                & 1                 & 6439.93                       &                                       & \\\hline
	2                & 1                 & 3268.65                       & 1.97$\times$                          & 98.5 \\\hline
	4                & 1                 & 1694.73                       & 1.93$\times$                          & 95.0 \\\hline
	8                & 2                 & 843.92                        & 2.01$\times$                          & 95.4 \\\hline
	16               & 4                 & 461.23                        & 1.83$\times$                          & 87.3 \\\hline
\end{tabular}
\caption{Scaling results for a single epoch of DDL-based training.}
\label{tab:ddlscaling}
\end{table}

\subsection{Comparing DDL and DDL With LMS}

Extending the model used in this work with LMS enabled an increase in input size
to 96$\times$96$\times$96 blocks, a scale which was not achievable without LMS.
The expectation from BP's team was that with a larger input, more of the
behaviors in input images would be captured without edge effects.
Table~\ref{tab:lms_accuracy} plots the change in accuracies between training
runs with and without the larger model, and illustrates small improvements in
accuracy for each class (particularly for class 1).
We also find that training the LMS-based model takes 13.7\% more time per
epoch.

\begin{table}
\centering
\begin{tabular}{| c | c | c | c | c |}
\hline
    \textbf{Model} & \textbf{Overall} & \textbf{Class 0}   & \textbf{Class 1} & \textbf{Class 2} \\\hline
    DDL            & 94.25\%          & 85.7\%             & 81.8\%           & 98.7\%           \\\hline
    DDL with LMS   & 94.78\%          & 86.6\%             & 84.4\%           & 98.9\%           \\\hline
\end{tabular}
\caption{Percent accuracies overall and for each individual class.}
\label{tab:lms_accuracy}
\end{table}

\subsection{Usability}

Perhaps the most significant result of moving from Vanilla Tensorflow to DDL was
the simplification of job creation, reduction in code size, and the improvement
in code maintainability. Whereas previous iterations of the same application
required complex generation of host configuration files and use of Tensorflow's
distributed APIs, the DDL APIs and job configuration are (subjectively) simpler
and easier to use. \texttt{ddlrun} is an \texttt{mpirun}-like utility for spawning multi-GPU DDL
jobs. Its support for basic sanity checks, its MPI-like usage, and its handling
of topology-aware configuration files yields a much better user experience than
spawning a Vanilla Tensorflow job. Additionally, DDL's integration with the Sun Grid Engine drastically simplifies
the application code required to support distribution across multiple GPUs.
Indeed, in the case of this test application, a simple \texttt{import ddl} was
all that was needed, allowing the deletion of dozens of lines of Vanilla
Tensorflow code. This is beneficial in the near-term for usability when building
new applications, and in the far-term in reducing the amount of code that needs
to be maintained. Using LMS was a similarly transparent experience, though there
is the potential need for more performance tuning with LMS.

\section{Summary}
A high bandwidth NVLINK connection between CPUs and GPUs has been exploited to train deep learning models that otherwise don't fit into the limited GPU memory. The high bandwidth link keeps the GPU utilization several times higher than the case where GPUs are connected to CPUs with  PCIe Gen3. The enabling tool, Large Model Support, can be used from TensorFlow and Caffe. Early user adoption has shown the applicability of LMS to practical deep learning work loads. LMS can be used in conjunction with distributed deep learning (DDL) to efficiently scale model training to multiple GPUs such that each GPU is utilizing the CPU memory for deeper models.

\section*{Acknowledgements}
The authors would like to thank Wladmir Frazao at IBM for his support that made this joint effort possible at all. Technical help on DDL from Brad Nemanich at IBM was invaluable in making this project a success. Also acknowledged are Charles Compton and Ian Watts from IBM for their help and support. 

\bibliography{learningsys}
\bibliographystyle{plain}

\end{document}